# Using a Model-driven Approach in Building a Provenance Framework for Tracking Policy-making Processes in Smart Cities

Barkha Javed*, Zaheer Khan, Richard McClatchey

University of the West of England, Bristol, UK

(Barkha.Javed; Zaheer2.Khan; Richard.Mcclatchey)@uwe.ac.uk

**Abstract**

The significance of provenance in various settings has emphasised its potential in the policy-making process for analytics in Smart Cities. At present, there exists no framework that can capture the provenance in a policy-making setting. This research therefore aims at defining a novel framework, namely, the Policy Cycle Provenance (PCP) Framework, to capture the provenance of the policy-making process. However, it is not straightforward to design the provenance framework due to a number of associated policy design challenges. The design challenges revealed the need for an adaptive system for tracking policies therefore a model-driven approach has been considered in designing the PCP framework. Also, suitability of a networking approach is proposed for designing workflows for tracking the policy-making process.

## 1 INTRODUCTION

The significance of policies is certain from its impact on the operation of cities and thus consequently on human life. For example, if a policy regarding control of air pollution is formulated then it will have impact on citizens' health and on development of cities (such as development of more green spaces, use of electric cars etc). Therefore, the role of policies is central in the smart cities endeavour [1, 2, 3]. However, policies are not developed merely by considering the problem or demand. Any policy is devised by taking into account all those concerns that can be influenced, directly or indirectly, by a newly created/modified policy. Furthermore, it is noted that policy-making is a political process whereby political agendas cannot be ignored; hence social, economic, and political agendas are all acknowledged while devising a policy. Similarly, constraints and conflicts in devised policy are also an important consideration. In summary the policy-making process needs to consider all associated factors, constraints, conflicts, and political agenda in order to serve both citizens and policy-makers. This suggests that any policy undergoes rigorous processing before being finalised and consumes and produces a large amount of information. This rich set of information provides insight regarding how a policy has been devised. This is similar to various settings such as business where data serves as an important asset and is used for analytics purposes; policy data can also be used for policy analytics.

Recent efforts [4, 5, 6] to explore policy analytics show its potential future. However, in order to enable and to explore potential analytics, detailed information regarding the execution of the process and decisions taken during policy-making process needs to be captured. This detailed information, in scientific terms, is called its provenance [7]. Such provenance captures all the details regarding a process and therefore provides a rich source of information for further usage. Consequently the aim of this research is to capture the provenance of policy-making processes which will not only provide the inherent benefits of provenance management [7, 8] but also a platform for analytics.

The complexities associated with a policy-making process however makes it quite challenging to track the process. This is because each policy is different and follows a different process and uses a diverse set of data for its creation; hence, a framework is required that can capture mutable requirements of the process and can adapt itself, thus enabling the provenance tracking of different policy scenarios. The inherent benefits such as reusability and dynamicity [9] of a model-driven approach show its suitability for addressing this challenge. Therefore, a model-driven approach has been considered for tracking the policy-making process and, as a consequence, meta-models and meta-meta-model for policy cycle are employed in this research.

A mechanism however is required to facilitate provenance capture; a workflow technology has been used for this purpose. However, workflow technology presents a number of challenges that are introduced due to the dynamic nature of policies, the involvement of diverse stakeholders, and the distributed setting. In order to address dynamicity, workflow details are also modelled in the meta-models. However, for facilitating diverse stakeholders' involvement, IP packet switching technique from computer networking based approach for workflows has been proposed.

This paper presents the Policy Cycle Provenance Framework (PCP Framework) which provides a foundation for tracking the policy-making process. To the best of our understanding and knowledge, the tracking of a policy-making process using a model-driven approach has not been hitherto tried and is therefore the main contribution in this research. Our second contribution is a networking based approach for workflow technology; this provides a mechanism to enable connection with diverse stakeholders and to address the distributed nature of policy-making process. It is important to mention that this is work in progress and reports design considerations for building the PCP Framework using model-driven approach. The final implementation and evaluation is out of scope of this paper and will be addressed elsewhere later.

The next section covers the literature review. Basis for PCP framework is elucidated in Section 3. Based on Section 3, PCP framework is presented in Section 4. Section 5 covers discussion and finally conclusion with future research direction is presented in Section 6.

## 2 LITERATURE REVIEW

### 2.1 Policy-making Process

A detailed literature review has been conducted to understand the policy-making process. It has been found that a number of researchers in the past couple of years have attempted to conceptualise the policy-making process [10, 11, 12, 13, 14, 15, 16, 17, 18, 19]. Of these the models defined by Macintosh [16], Khan et al [17], and Sonntagbauer et al [18] have been conceived for the generic policy-making cycle (shown in Figure 1). A generic policy-makiing cycle for devising policies has also been studied in detail by Tsohou et al [19]; they analysed twenty five policy making processes from four countries: UK, Hungary, Portugal, and Turkey. Their analysis clearly demonstrates that the policy making process can be considered a cyclic process as identified by Macintosh, Khan et al, and Sonntagbauer et al [16, 17, 18].

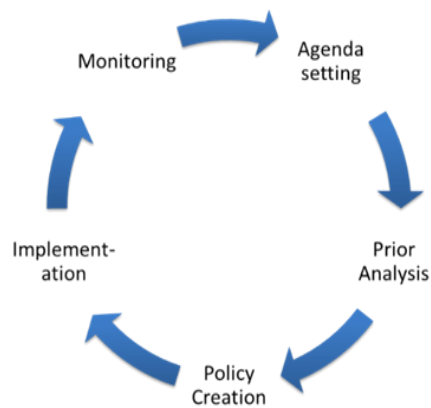

**Figure 1: Policy Cycle**

The policy-making process employs a large source of information for formulating a policy. These details when captured can provide a valuable insight regarding the process followed and the data produced/consumed for devising a policy. Therefore, provenance has been considered for capturing the policy-making process and to provide a platform for analytics.

**Note:** This paper defines 'agenda setting', 'prior analysis', 'policy creation', 'implementation', and 'monitoring' as *phases or stages* of a policy-making cycle. The activities involved in each phase are called *tasks*.

### 2.2 Provenance

Information regarding the origin/derivation, ownership and usage of data is important for data authenticity and verification. Such information is often termed provenance data [7]. It can refer to the history of artifacts, e.g. the history of the ownership of a painting or the steps involved in conducting a scientific experiment where experimental data and processes are required for re-computation, verification, and repetition of experiments [20]. Provenance captures the what, how, who, which, when, where, and why data of object usage. Such information provides details regarding process execution. The capture of the policy-making process can benefit from the inherent characteristics of provenance data such as data quality, verification and validation of information, reliability and accuracy, integrity, authenticity, validation of attribution of data, and transparency and trust in the system. Furthermore, provenance provides a rich source of information which can be used to perform analytics [21, 22, 23].

**2.3 Provenance in Policy-making Process**

The need for tracking the policy-making process has stimulated considerable research. A number of case studies were conducted by Sajjad [24] in order to uncover the suitability of workflow technologies for the policy-making process. The case studies clearly demonstrate the potential of provenance for the policy-making process. Edwards. P et al., [25] show the need for provenance in evidence-based policy making and track the process of evidence creation for policies; evidence used within policy work is captured in order to provide an audit trail for the system. Policies are modeled by Scherer. S et al., [26]; the processes being followed to model policies are tracked in order to provide evidence of how policy simulations were created.

*Analysis*

The implication of provenance for policy-making is clear from existing researches [24, 25, 26]. However, it has been found from the research that at present provenance has been considered for either simulation purposes or for capturing creation of evidence that is to be used in the policy-making process. However, it has not been found that how provenance can be used to track the policy-making process i.e. track when a policy is being devised. Therefore, to enable tracking of a policy-making process (shown in Figure 1) this paper introduces a provenance framework.

**3   BASIS OF POLICY CYCLE PROVENANCE FRAMEWORK (PCP Framework)**

To unearth the basis of a PCP framework, an in-depth investigation of the process has been carried out. Four case studies from [19] have been employed to uncover the details required for designing the provenance framework. The case studies are not explained in this paper as it has already been covered in detail in [19]; this paper only discusses the analysis of the case studies.

The case studies covered by Tsohou et al [19] revealed some important observations, considerations of which are vital for conceptualising the provenance framework for the policy-making process. Case studies 1, 2, and 3 show that at any phase of the policy-making cycle, a loop back to the previous phase(s) is required. It is not necessary that any phase of the policy-making process is only associated with the next phase.

The authors have specified tasks associated with each phase of the policy-making cycle. However, the case studies show that a clear demarcation between stages does not exist. This makes it quite challenging when identifying which tasks are associated with each phase. This also makes it difficult to understand how Tsohou et al [19] identified the different stages of the policy-making process in the four case studies. Therefore, further literature [16, 17, 18] was studied to uncover tasks for each phase of the process (the tasks are shown in Table 1). This is required to clearly define which tasks each phase is required to execute. Defining high-level or mandatory tasks for each phase will assist to structure the process and in designing a provenance framework. Defining mandatory tasks however would not be fruitful if it is not known which details are covered by the identified tasks for each phase of the policy-making cycle. This is required because it will assist in defining a generic description of tasks to which diverse set of policies can apply. Consequently, detailed tasks have been identified from [16, 17, 18].

Case studies also show that a chronological order may not exist between tasks in a phase i.e. it is not necessary that all tasks (as shown in Table 1) of any phase are executed in a well-defined sequence. There also exists a possibility that some tasks may or may not require execution depending on policy requirements and human decision. However, there still exists some rules and regulations regarding the execution such as *implementation phase* cannot be executed before *agenda setting phase*. Similarly, in *Agenda Setting phase*, *'validation'* task cannot execute before *'problem identification'* task. However, after *'problem identification' task* either *'validation'* can be executed or *'plan setting'* depending on the policy requirements.

Uneven patterns regarding citizens' participation have been observed in four case studies. Case study 1 involves citizens in 'agenda setting', 'prior analysis', and 'monitoring' phases. Case study 2 involves citizens in 'policy creation' and 'policy implementation' phases. Case studies 3 and 4 involve citizens in only 'agenda setting' phase. The process indicates that current practices [28, 29] in policy making incorporate new governance models which include a top-down as well as a bottom-up approach and therefore it is an important consideration for the provenance framework.

Table 1: General Tasks [16, 17, 18, 19]

| Policy Lifecycle | General Tasks |
| --- | --- |
| Agenda Setting | i) Problem Identification<br>a) acquisition of qualitative and/or quantitative data<br>b) review of collected data/reported issue<br><br>ii) Validation<br>a) evidence gathering for objective or subjective validation<br>b) analysis of gathered evidence |

| | iii) Plan setting<br>a) Identify action to be taken (change of existing policy or devise new policy)<br>b) devise strategy |
|---|---|
| Analysis | i) Challenges and opportunities identification<br>a) specification of goals<br>b) data collection from diverse sources<br>c) collection of opinions from stakeholders<br>d) analysis of collected data<br><br>ii) Determination of solution approaches and strategies<br>a) develop a range of options<br>b) analysis of options |
| Policy Creation | i) Formal Consultation<br>a) collection of residents' opinions<br>b) stakeholders' engagement<br>c) assessment of opinions<br><br>ii) Final Decision and approval<br>a) weighing of policy options in the political context<br>b) decision based on step 'a'<br><br>iii) Policy Formulation<br>a) draft policy based on policy options<br><br>iv) Design implementation and monitoring plan<br>a)Actions to be taken for implementation and monitoring |
| Policy Implementation | i) Interagency collaboration<br>a) collection of data<br>b) selection of relevant implementation body<br><br>ii) development of regulation/legislation<br><br>iii) Collection of data – also called monitoring data<br>a) identify key indicators of monitoring |
| Policy Monitoring and Evaluation | i) Monitoring<br>a) collect evidence<br>b) analyse data collected as per specified indicators<br>c) collect views/feedback of users including citizens<br>d) analyse collected views<br><br>ii) Evaluation<br>a) Administrative and judicial evaluation<br>b) Impact evaluation<br><br>iii) Loop back to stage one |

All case studies revealed that it is not necessary that every task in each phase must be executed by the same department. Different tasks can be executed by different departments which raises issues regarding tracking, maintaining, and integrating data at one place.

Observations identified from the case studies are summarised in Table 2.

**Table 2: Observations drawn from the Case Studies [19]**

| Sr. No | Observations [13] | Policy Framework Design Needs |
|---|---|---|
| 1 | No chronological order for the execution of tasks | A mechanism is required that enable smooth flow of process |

| 2 | The policy-making process is cyclic. However, it has been analysed that any step can loop back to previous stage(s) depending on the problem/requirement. | Connection is not required only with the subsequent phases but also with previous phases |
|---|---|---|
| 3 | Every case study followed different steps at each phase of the process i.e. the process followed by policy-making varies from one problem to another. Therefore, data used and produced at each phase of the process is also variable. | An approach is required that must be adaptable |
| 4 | Uneven pattern of citizens' participation in policy-making process. | System must be adaptable in order to facilitate this observation |
| 5 | No clear demarcation between phases of the policy-making process. | Tasks for each phase needs to be identified |
| 6 | Same task can be executed by different departments therefore data governance is another challenge. | Out of scope of this research |

As observed from the analysis of the case studies, the policy-making process involves diverse stakeholders and different departments which introduces issues such as data governance and provenance security. Though this an important consideration in a real-time setup but it is not within the scope of this research.

Based on the observations, policy framework design needs have been identified (as shown in table 2). Observations 1 and 2 show that for the collection of provenance, flows in the policy-making process need to be specified. Therefore, workflows are established for orchestrating the policy-making process. The tasks identified in Table 1 (as a result of observation 5) serve as a foundation for structuring the process and specifying the connections among tasks and phases. These tasks will also help to specify constraints in the system. From observations 4 and 5, it has been identified that each policy is different thus requiring different process and data requirements. Considering this, a model-driven approach has been considered as a viable choice for tracking the policy-making process; using this approach changes can be accommodated in the provenance framework.

A model-driven approach has been used in OCOPOMO [26] for agent-based policy modelling. In this project, stakeholders identify the issues in an existing or new policy and specify scenarios. These scenarios are employed to generate different policy models which provides an insight of the impact of policies. During the process of policy simulation, the provenance is tracked and traces/links of simulation with scenarios and evidence are captured. This provides traceability and accountability in the policy modelling-process.

The current work [26] shows the applicability of the model-driven approach in the policy domain but it has not been used to enable the tracking of policy creation (i.e. when a policy is being devised). A model-driven approach for tracking the complete policy-making process (shown in Figure 1) has not been tried before. A provenance framework for policy cycle thus takes this into account. Furthermore, a model-driven approach will also assist in handling the connection among tasks and phases ( addressing observations 1 and 2) and in specifying constraints regarding the process flow. If any change in the process is required, then changes in the meta-models can be accommodated thus automatically reflecting changes in the system. Furthermore, meta-models help to define a generic provenance framework and therefore are suitable for capturing provenance of diverse policies.

## 4 PCP Framework

From the observations and analysis in section 3, the PCP framework has been designed accordingly and is shown in Figure 2.

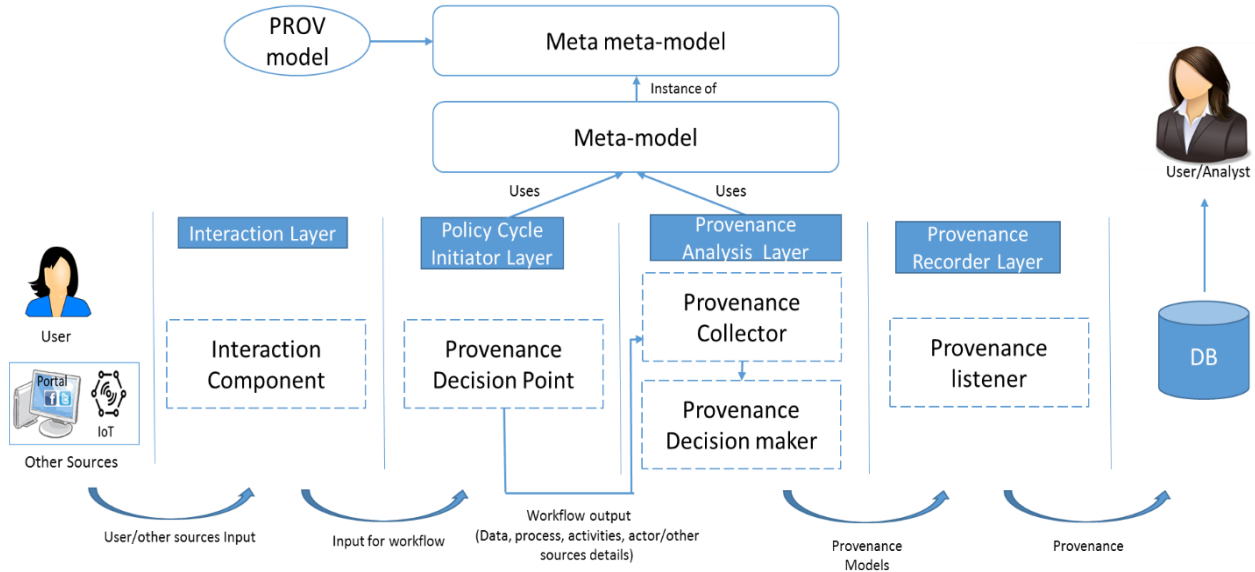

Figure 2: PCP Framework

### 4.1 Working of PCP Framework

Figure 2 shows the PCP framework. This framework has four layers: Interaction, Policy Cycle Initiator, Provenance Analysis, and Provenance Recorder Layer. These layers ensure separation of concerns which confine related logic to particular layers; therefore any change in the layer such as delete or addition of new components can be easily accommodated. The *Interaction Component* in *Interaction Layer* is responsible for taking input from the users or from other sources. This input is passed to the *Provenance Decision Point* component in the *Policy Cycle Initiator* layer. This component encapsulates the design of underlying workflow needed for policy-making cycle. However, unlike scientific workflows, the operation of policy-making workflow will be largely human-centric (but there would also be input from other sources) and therefore regular interaction with users is required. Furthermore, the user interaction is not restricted for starting the first phase/task of the policy cycle. The user input is also required for many intermediate tasks and thus regular interaction is required. Figure 2 also shows that the *Policy Cycle Initiator* uses a meta-model. This is because the meta-model of the policy cycle also specifies structure, process/data flow, and constraints. Considering the constraints, pre-conditions, and post-conditions, the process flow (and also data flow) will be orchestrated by the workflow. The provenance data produced include information such as process, associated data, events, artefacts, actor/user details, activities executed, workflow details, comments, decisions, etc. The generated activities are passed as XML or JSON documents to the *Provenance Analysis Layer.*

The *Provenance Analysis Layer* also uses a meta-model. It takes the provenance data and, according to the meta-models, it generates provenance models. The meta-models ensure that the output from the *Policy Initiator Layer* can be handled by the *Provenance Analysis Layer*. The output models from the *Provenance Analysis Layer* are passed to the *Provenance Recorder Layer* who is responsible for storing provenance in a database. Different storage options are available for the storing provenance including relational database, XML, JSON, graph database. A relational database is not a viable option due to the associated complexity in capturing the relationships among provenance data. Therefore, XML or graph databases are suitable design choices; we are currently evaluating the suitability of these databases.

The stored provenance can be used by analysts for analytics purposes. However, we are currently exploring that what possible analytics could be performed on the policy-making provenance.

**4.2 Design Considerations in the Use of Meta-models and Workflows**

This section is dedicated to elaborating the detail design considerations of workflows and meta-models in the PCP framework.

**4.2.1 Meta-models**

From the generic policy-making cycle (i.e. figure 1), there are two possible design consideration as shown in Figure 3.

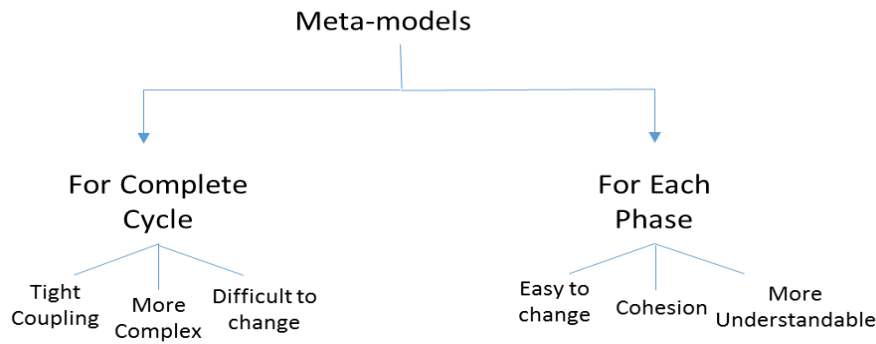

**Figure 3: Design Consideration for Meta-models**

Figure 3 displays the design consideration for the use of meta-models for the policy cycle. Each phase of the process encapsulates a number of tasks (as presented in Table 1); defining a meta-model for the full policy-making cycle (i.e. covering all phases) is possible but will add to the complexity of the system. It makes it cumbersome to make any change in the meta-model. Furthermore, since it covers the full lifecycle therefore there will be tight coupling between stages. On the other hand the meta-model for each phase lowers the complications; this will also assist to confine changes and will ensure cohesion.

Another reason for not defining a meta-model for the full lifecycle is the challenge allied to defining at the outset the association of a particular task in a phase with a particular task in the subsequent phase. The reason behind this is the uncertainty that is associated with the execution of activities in any phase (from observation of the case studies). In order to understand this, suppose that **phase 1** has three tasks and **phase 2** has five tasks. It is not mandatory that all tasks in a phase will be executed for a policy. Consider **Policy A** that executed tasks 1, and 3 which suggests that task 3 must be associated with next phase. Here also task 3 of phase 1 can be linked to any task of any phase (not necessarily task 1 of phase 2). This clearly indicates uncertainty in the policy-making and is thus an important consideration when deciding a design of the system. The challenge of defining pre-defined connections is depicted in Figure 4.

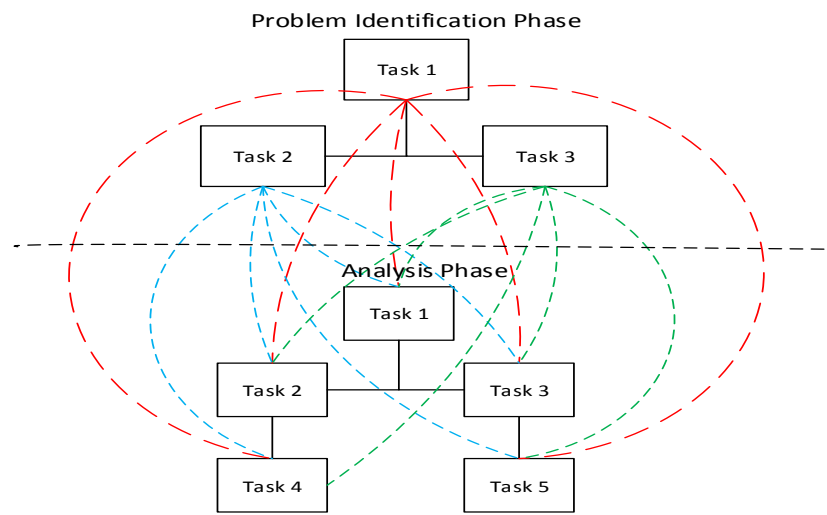

**Figure 4: Challenge of specifying connection between stages**

Considering the aforementioned challenges, a meta-model <u>for each phase</u> of a policy cycle has been considered.

Figure 1, however, clearly shows that the connection between stages of a policy cycle is required which signifies that another model is required that must specify the connection between stages. Therefore, a meta meta-model for a full policy cycle is considered. The meta meta-model defines the meta-models for each phase and is also self-descriptive.

The provenance of policies is not only required to capture the process but also the data that is being consumed, produced, and transformed during the process. Furthermore, it is also important to track details of actors involved in the process because policy-making is largely a human-centric process. These details suggest that meta-models (and meta-meta-models) need to consider structure, process, data, actors, activities, artefacts, decisions, constraints, workflow, and the connection among these entities. Our current analysis shows that the PROV model [28] can be used for meta-models and meta-meta-models. However, we are currently investigating in detail which features of the PROV model would be suitable and also exploring what extensions to the PROV model are required for capturing the provenance of the policy-making process.

### 4.2.2 Workflows
#### 4.2.2.1 Connectors
Workflows have been considered in order to enable tracking of a policy-making process. For each stage a separate workflow (we call it 'inner-workflow') is defined. However, for the connection between stages another workflow (we call it outer-workflow) is required that should cover the complete policy cycle; nevertheless the complexity of connections between stages is evident from Figure 4. Therefore, an approach is required that can facilitate the smooth flow of information between stages. Table 2 and Figure 4 clearly show that a chronological order may not exist between the tasks in a phase. This further raises the challenge of how tasks (of a phase) would be connected with other stages. In order to address this issue, a connector is specified for each phase of the process with which an internal-workflow is connected. This approach has been employed because it will easily accommodate change in a process at run-time. For example, if the *agenda setting* phase has 'n' number of possible activities then there is a need to identify a specific end task for it to be connected to the next phase. In this case, the first activity of the *analysis phase* is also required to be identified. The policy-making process, however, is not without uncertainty. The specification of any pre-defined connections between stages is not a feasible approach. Therefore, when an internal-workflow is to be executed, the details of the last activity will be stored in a connector. The connector in one phase will communicate with the connector in the next phase. The connector in the next phase will determine, based on user interaction or data, the activity with which the previous phase needs a connection. This means that the proposed connectors will also act as decision making nodes to enable communication between different stages and tasks. This is important because provenance must capture the process flow (which also encapsulates data flow) in order to have a complete view regarding the policy process.

This will also help to address observation 2 (as given in Table 2). A loop back to the previous stage is also uncertain; it cannot be predicted beforehand with which tasks a policy-making phase needs to be connected. Therefore, this uncertainty makes it difficult to identify all possible connections. Furthermore, this approach will also facilitate run-time decision to establish connections between stages.

#### 4.2.2.2 Using Networking Approach for Policy Cycle Workflows
The case studies revealed the involvement of a large number of stakeholders, various departments, and various consultees (departments not co-located) at each stage of the policy-making process. This distributed setting further exacerbates the challenge of tracking the policy-making process since it requires a workflow to span all departments/consultees. In order to address this, a concept similar to *'Internet Protocol packet-switching communication in computer networks'* has been considered. Each phase holds the identity addresses of different departments/stakeholders (including consultees); whenever there is a requirement for the collection or feedback/analysis, then based on this address, the request is forwarded to the concerned department. We call these requests *'tokens'* which carry information regarding the required details from the stakeholder (s). This has been achieved by using an *'external connector'* (similar to router) which holds the addresses of various bodies. Using the *external connector* and *token*, the receiving stakeholder provides the detailed information which will be passed to the *external connector* who will then be responsible for giving the input to a workflow. Provenance information will be collected at those points when a request is passed to the *external connector* or any input is received from an *external connector*.

Each task of a phase is associated with an external connector because it is not necessary that a specific task of any phase is executed solely by the same department or by same actor. For example, if we consider the *agenda setting* phase then suppose an analysis in a *problem identification* task is required from some external consultee or from another actor. In such a case the workflow is required to be connected with workflows of all external consultees which is a very challenging practical scenario. Therefore, by using an *external connector*, the policy department forwards the request to the external consultee. The next activity of a policy-making process will not executed until and unless a response is received from them. The response will be treated as input for the next stage or for task of any stage.

The design approach is shown in Figure 5 (however only first three phases are shown in the figure but the connectors and external connectors will be specified for remaining phases too. This Figure only demonstrates the approach).

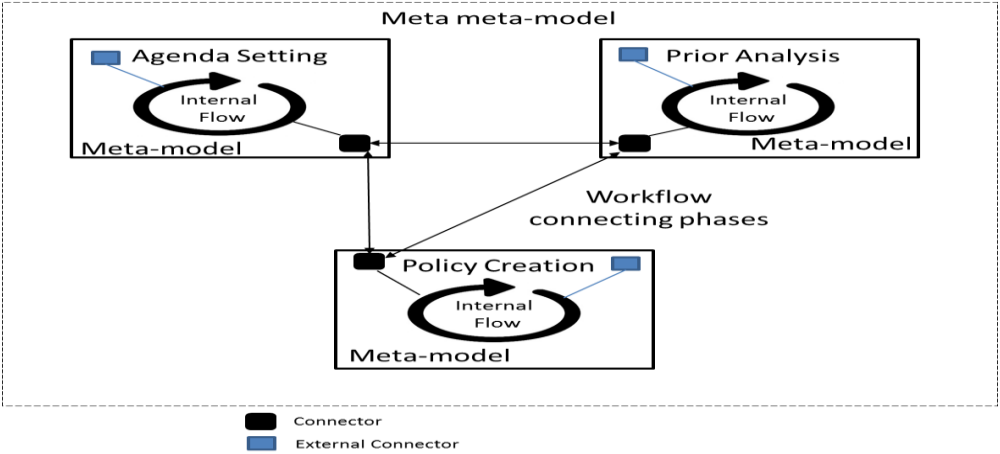

**Figure 5: Design Approach**

## 5. Discussion

The PCP framework given in section 4 shows the suitability of a model-driven and workflow approach for capturing provenance in a policy cycle. The PCP framework has been conceived in a manner that generality can be ensured. This means that the meta-models and meta-meta-model control changes in the system; if any change is required then user can play with meta-models without having to go into the details of other components. The use of meta-models will help to define varying scenarios of policies thus enabling capturing of different policies. Meta-modelling approach for tracking the policy cycle has not been considered before and will result in a novel contribution. The need for workflows for policy making has been argued by Sajjad [24]; the author presented a framework for workflow **technology adoption** in policy domain but its design and implementation has not been covered and researched on. We have therefore proposed to adopt the IP based packet switching approach from computer networks domain for addressing complexities associated with tracking the policy-making process. This will be investigated in detail as part of future research work.

Due to integrated nature of policy-making process, there may arise issues related to data security and privacy. But these issues are not considered in this paper and are out of the scope of this research.

## 6. Conclusion

This paper proposed a generic provenance framework for the policy-making process (PCP) that can be used for different policies setups and serve as a platform for analytics. In order to define a provenance framework, an extensive literature review was carried out and based on identified case studies, a number of observation were drawn. These observations exhibit the suitability of a model-driven approach for the implementation of a policy cycle provenance framework. A model-driven approach has therefore been used for developing an adaptive system. Furthermore, the creation of a policy requires a process and therefore a set of activities to be executed in order to conceive any policy. Thus orchestration of the policy-making process is required; workflows have been evaluated as an appropriate option to this end. The identified challenges of using workflows in a policy-making process assisted in designing a networking approach for workflow technology. In this paper we only introduced PCP framework and discussed the design consideration. For future work, this research is intended to develop a provenance system as a proof of concept which will be used for validation of the proposed approach.